\documentstyle[12pt]{article}
\def\a{\alpha}
\def\b{\beta}

\def\tM{\tilde{M}}
\def\tt{\tilde{\tau}}
\def\tr{\tilde{r}}

\def\P{\Phi}

\def\t{\tau}

\def\Z{{\bf Z}}

\begin{document}
 
\newcommand{\inv}[1]{{#1}^{-1}} 
\renewcommand{\Re}{{\rm Re \,}}
\renewcommand{\Im}{{\rm Im \,}} 
\renewcommand{\theequation}{\thesection.\arabic{equation}}
\newcommand{\beq}{\begin{equation}}
\newcommand{\eeq}[1]{\label{#1}\end{equation}}
\newcommand{\ber}{\begin{eqnarray}}
\newcommand{\eer}[1]{\label{#1}\end{eqnarray}}
\begin{titlepage}
\begin{center}

                                \hfill    CERN-TH/96-127 \\
                                \hfill    NYU-TH-96/05/01\\
                                \hfill    hep-th/9605118\\

\vskip .5in
 
{\large \bf Aspects of Space-Time Dualities}\\

\vskip .4in
                          
{\large  Amit Giveon}\footnotemark \\
\footnotetext{On leave of absence from the 
Racah Institute of Physics, The
Hebrew University, Jerusalem 91904, Israel; e-mail address: 
giveon@vxcern.cern.ch}
\vskip .1in
{\em  Theory Division, CERN, CH-1211, Geneva 23, Switzerland} \\
\vskip .15in
and
\vskip .15in
{\large  Massimo Porrati}\footnotemark \\
\footnotetext{e-mail address: porrati@mafalda.nyu.edu}
\vskip .1in
{\em Dept. of Physics, NYU, 4 Washington Pl., New York NY 10003, USA}
\vskip .1in
 
\end{center}
\vskip .2in
\begin{center} {\bf ABSTRACT } \end{center}
\begin{quotation}
\noindent
Duality groups of Abelian gauge theories on four manifolds and their
reduction to two dimensions are considered. The
duality groups include elements that relate different space-times in 
addition to relating different gauge-coupling matrices. We interpret (some of)
such dualities as the geometrical symmetries of compactified theories in higher
dimensions. In particular, we consider compactifications of a 
(self-dual) 2-form in $6-D$, and compactifications of a self-dual
4-form in $10-D$. Relations with a self-dual superstring in $6-D$ and with the
type IIB superstring are discussed.

\end{quotation}
 
\vfill

\begin{flushleft}
CERN-TH/96-127\\
May 1996
\end{flushleft}
 
\end{titlepage}
\eject
\def\baselinestretch{1.2}
\baselineskip 16 pt

\noindent
\section{Introduction}
\setcounter{equation}{0}

Electric-magnetic dualities in gauge theories and string duality symmetries 
have recently been studied extensively (for a review, see for instance
\cite{ggpz3,se,gpr,p} and references therein).
Sometimes, by using string dualities, electric-magnetic duality can be 
related to a geometrical symmetry of the internal space in some string
compactification. 

In this work, we consider duality symmetries in $4-D$, Abelian 
gauge theories which involve also the (Euclidean, compact) 
space-time $M^4$. Such dualities -- rather mysterious from the $4-D$ point of
view -- are better understood as geometrical symmetries of theories in higher
dimensions, compactified to $M^4$ on some internal space.

Explicitly, we find dualities which relate a pair $(M^4,\t)$ to a different
pair $(\tM^4,\tt)$, where $\tau$ is the (complex) coupling constant matrix of a
$U(1)^r$ gauge theory on $M^4$, and $\tt$ is the dual coupling constant matrix
of a $U(1)^{\tr}$ gauge theory on $\tM^4$. In general, $M^4$ and $\tM^4$ have
not only a different geometry, but also a different topology; in this case,
$\tr\neq r$.\footnote{In this case, the duality is expected to be only
a symmetry of the classical part of the theory.}

Some of the dualities considered here can be understood as the consequence of
string dualities, in the limit where gravity is decoupled.

Let us start with a simple example. String-string triality  relates the
heterotic string compactified on $T^4\times T^2$ to type IIA and type IIB
strings compactified on $K^3\times T^2$. The duality group of the heterotic
string includes, in particular, the $SL(2,\Z)_S\times SL(2,\Z)_U\times 
SL(2,\Z)_T$ acting on the dilaton-axion field $S$, the complex structure of 
the two-torus $U$ and its complex K\" ahler structure $T$, respectively,
as well as the
$\Z_2$ factorized duality (mirror symmetry) interchanging 
$U\leftrightarrow T$ \cite{gpr}. String-string duality involves also
the $\Z_2$ interchanging $U\leftrightarrow S$ or $T\leftrightarrow S$.
The low-energy effective field theory, in the infinite Planck-mass limit, is 
an $N=4$ supersymmetric Yang-Mills (YM) 
theory on Minkowski space. At a generic point in moduli space, it is an
Abelian gauge theory, including a $U(1)^4$ gauge group originating from the
internal torus $T^2$. One expects the dualities described above to be manifest
in the $N=4$ YM theory. Indeed, the $SL(2,\Z)^3$ and $\Z_2$'s, corresponding to 
$U\leftrightarrow T$, $U\leftrightarrow S$ and $T\leftrightarrow S$, are part 
of the $Sp(8,\Z)$ duality transformations acting on the $4\times 4$ gauge 
coupling matrix $\t$ of the $U(1)^4$ gauge theory.

Our second example, described in section 2, involves also space-time. We argue
that an $SU(2)$, $N=4$ YM theory, broken to $U(1)$ at large scalar VEVs, is
invariant under an $O(2,2,\Z)$ duality group acting on the complex gauge
coupling $S$ and the complex structure $U$ of two space directions. This
duality group includes the well known $SL(2,\Z)_S$ $S$-duality group, the
geometrical  $SL(2,\Z)_U$ symmetries and, in addition, a duality which  
interchanges the gauge coupling with the complex structure of space: 
$S\leftrightarrow U$. In section 2, we also explain the origin of this
duality from the manifest geometrical symmetry of a $6-D$ self-dual
superstring, compactified to $4-D$ on $T^2$.

In section 3, we consider a compactified theory of a 
(self-dual) 2-form in $6-D$ 
and show that its geometrical symmetries lead to the dualities of section 2, as
well as several other dualities appearing in string theory. 

In section 4, we discuss more dualities in Abelian gauge theories, which
relate different pairs of $(M^4,\t)$. Moreover, we show how such dualities are
a consequence of geometrical symmetries in a theory of self-dual 4-form in
$10-D$, compactified on $M^4\times T^2 \times \tM^4$.   

\noindent
\section{$S\leftrightarrow U$ Duality in $N=4$ Yang-Mills Theory and the
Self-Dual Superstring in $6-D$}
\setcounter{equation}{0}

Our second example involves a compact space-time. Consider an $SU(2)$, $N=4$ YM
theory on $M^4=S^1_{\b}\times S^1_R\times T^2_U$. Here $S^1_{\b}$ is a compact
time at radius $\b$ (= the inverse temperature), $S^1_R$ is a compact space
coordinate on a circle with radius $R$, and the other two space coordinates are
compactified on a two-torus $T^2_U$ with complex structure $U=U_1+iU_2$. 
We consider the
limit in which $\b, R\to \infty$ such that their ratio is finite, say $\b/R=1$.
At large vacuum expectation values of the scalar fields $\langle \P\rangle \to
\infty$, a simple generalization of the computation in \cite{ggpz} gives the
partition function
\beq
Z(S,U)=c\sum_{n} \exp\{-\pi n^t M(U)\otimes M(S) n \} .
\eeq{ZUS}
Here $c$ is an $S$--independent and $U$--independent factor, $n$ is a 4-vector
of integers, $S=S_1+iS_2=\theta/2\pi +i4\pi/g^2$ 
is the complex gauge coupling and
\beq
M(S)={1\over S_2}
\left(\begin{array}{cc} 1 & S_1 \\  S_1 & |S|^2
\end{array}\right) .
\eeq{MS}
The $2\times 2$ matrix $M(U)$ is given by (\ref{MS}), with $S$ replaced by
$U$.

The partition function in (\ref{ZUS}) is invariant under the $O(2,2,\Z)$
duality group acting on $U,S$ \cite{gpr}. This duality group is the product 
of $SL(2,\Z)_S$, $SL(2,\Z)_U$ and the $\Z_2$ interchanging them.
The $SL(2,\Z)_S$ is the well-known $S$-duality group acting on $S$.
The $SL(2,\Z)_U$ is the geometrical symmetry of the torus $T^2_U$, acting on
its complex structure $U$. The rather surprising duality transformation is the
one interchanging the gauge coupling with the structure of space
$S\leftrightarrow U$! It is a mysterious symmetry from the $4-D$ gauge theory
point of view.

We can understand the origin of this duality from the manifest geometrical
symmetry of a self-dual superstring in $6-D$ \cite{w,s} compactified on 
$S^1\times S^1\times T^2\times T^2$. This string theory is non-gravitational,
defined in rigid $6-D$ space-time, and is supposed to give a theory with chiral
$N=2$ supersymmetry in $6-D$. 
The string is coupled to a self-dual 2-form $B$ (namely, $B$ with a 
self-dual field strength  $H$: $H=dB=*H$ in Minkowski space, $*H=iH$ in
Euclidean space). It was argued that compactifying
this theory to $4-D$ on $M^4\times T^2$ gives rise to an $N=4$ 
supersymmetric gauge theory \cite{w,s}; the $B$ field gives rise to a 
$U(1)$ gauge field in four dimensions. The winding numbers of the string 
around the two
independent cycles of the torus correspond to electric and magnetic charges. 
The $S$-duality symmetry of the $4-D$, $N=4$ theory is a simple consequence 
of the geometrical symmetry of the torus.
In the decompactification limit, all the charged states become
infinitely massive, and one is left with a free Abelian theory in four
dimensions.

Consider first the $6-D$ self-dual superstring compactified to $4-D$ on
$T^2_S$, and with $M^4=S^1_{\b}\times S^1_R \times T^2_U$. Recall that the
subscripts $S$ and $U$ denote the complex structures of $T^2_S$ and $T^2_U$,
respectively. At the decompactification limit, the partition function of the
theory is $Z(S,U)$. 

The $S\leftrightarrow U$ duality is manifest if we consider instead the
compactification of the $6-D$ theory to $4-D$ on $T^2_U$, and with 
$\tM^4=S^1_{\b}\times S^1_R \times T^2_S$.
Again,  at the decompactification limit, the theory is a free 
$D=4$, $N=4$ supersymmetric $U(1)$ gauge theory, but this time with a gauge
coupling $U$, and a complex structure $S$ in space-time. 
Therefore, the partition function of the theory is $Z(U,S)$.

Since both theories are the same $6-D$ self-dual theory, we must find
\beq
Z(S,U)=Z(U,S) .
\eeq{ZZ}
This duality indeed obtains in the partition function  eq.~(\ref{ZUS}),
but the argument given before has a subtlety: no manifestly
Lorentz-invariant action for a self-dual two-form exists. Thus, while the
$S\leftrightarrow U$ duality of the equations of motion of an Abelian $4-D$
gauge theory can be deduced along the previous lines, 
the duality of the partition function itself must be proved in another way.
This problem is studied in the next section.

\noindent
\section{Compactifications of 2-Form Theories in $6-D$ and Duality}
\setcounter{equation}{0}
Let us consider in more detail the result of the previous section. We want to
compactify a $6-D$ two-form on the manifold $T^2_T\times T^2_U\times T^2_S$,
and impose an appropriate condition of self-duality. In the previous section
the modular parameter $T$ was $T=i\b/R$. 

The six-dimensional action is~\cite{v} 
\beq
S_{6-D}={1\over 2\pi i}\int dB \wedge H - {1\over 4\pi}\int *H \wedge H . 
\eeq{m1}
We denote by $a_S$ and $b_S$ the cycles of the torus $T_S^2$, obeying
\beq
\int_{T^2}a_S \wedge *a_S={|S|^2\over S_2},\;\;
\int_{T^2}a_S \wedge *b_S={S_1\over S_2},\;\;
\int_{T^2}b_S \wedge *b_S={1\over S_2},\;\; S=S_1+iS_2.
\eeq{m2}
The self-duality condition is imposed as in~\cite{v,ht} by choosing a global
vector field $V$ in $T^2_S$ such that $i_Vb_S=0$, and 
setting half of the components of the field-strength $H$ equal to $dB$:
\beq
i_V(H-dB)=0.
\eeq{m3}  
The resulting $4-D$ action is an Abelian gauge theory with
coupling $S$~\cite{v}. Compactifying its action on $T_U^2$ gives rise to 
a $2-D$ sigma model. Obviously, we could have reduced the $6-D$ theory 
directly to $2-D$ on $T_S^2\times T_U^2$. Indeed, the reduction was 
performed
in~\cite{v} in the case of a generic four-manifold $M^4$ with equal number
of self-dual and anti-self-dual two-forms ($b^+=b^-$). 
Following~\cite{v}, we denote by $\alpha_i, \beta^i$, $i=1,..,b^+$ a 
basis of $H_2(M^4,\Z)$ such that
$\int_{M^4}\alpha_i\wedge \beta^j=\delta_i^j$, and we define
\beq
\int_{M^4}\b^i\wedge *\b^j=G^{ij},\;\;
\int_{M^4}\a_i\wedge *\b^j=B_{il}G^{lj},\;\;
\int_{M^4}\a_i\wedge *\a_j=G_{ij}-B_{il}G^{lm}B_{mj}.
\eeq{m4}
The self-duality condition is always given as in eq.~(\ref{m3}), with the
global vector $V$ obeying $i_V\b^i=0$ for all $i$.
The $2-D$ sigma model obtained by reducing eq.~(\ref{m1}) on such manifold
is~\cite{v}
\beq
S_{2-D}={1\over 2\pi} \int dX^i\wedge (G_{ij} *dX^j -iB_{ij}dX^j).
\eeq{m5}
This is the action of a sigma model on the toroidal background with metric
$G_{ij}$ and antisymmetric tensor $B_{ij}$.
 
In our case $M^4=T^2_S \times T^2_U$, and a basis of $H_2(M^4,\Z)$ is
\ber
&& \alpha=a_S\wedge a_U,\;\; a_S\wedge b_U,\;\; a_S\wedge b_S, \nonumber
\\ &&
\beta=b_U \wedge b_S,\;\; b_S\wedge a_U,\;\; a_U\wedge b_U.
\eer{m6}  
The vector $V$ is the same as before: it is oriented along a cycle of 
$T^2_S$ and obeys $i_Vb_S=0$.
The metric $G_{ij}$ and antisymmetric tensor $B_{ij}$ 
describe the space $T^2\times S^1$, where $T^2$ is a torus with complex
structure $U$ and complex K\"ahler modulus $S$, while the radius of $S^1$
is $\sqrt{V_U/V_S}$, where $V_U$ ($V_S$) is the volume of 
$T^2_U$ ($T^2_S$): 
\beq
G_{ij}=\left( \begin{array}{ccc} {S_2|U|^2/U_2} & {S_2 U_1/U_2} & 0 \\ 
{S_2 U_1/ U_2} & {S_2/U_2} & 0 \\
0 & 0 & V_U/V_S \end{array} \right), \;\;
B_{ij}=\left(\begin{array}{ccc} 0 & S_1 & 0 \\ -S_1 & 0 & 0 \\ 0 & 0 & 0
\end{array} \right).
\eeq{m7} 
Notice that $\sqrt{V_S/V_U}$ is the coupling constant of the sigma model on 
$S^1$; thus, in the limit $V_S/V_U\rightarrow \infty$, $S^1$ decouples and
one is left with a $2-D$ sigma model on $T^2$. The $S^1$ sigma model
possesses a $T$-duality which inverts the radius. Thus, $S^1$ decouples also
in the limit $V_S/V_U\rightarrow 0$. On $T^2$, the interchange
$S\leftrightarrow U$ is simply the mirror symmetry of the 
torus~\footnote{The $S\leftrightarrow U$ duality holds for any $V_U, V_S$; 
but for triality, considered below, one should take $V_S\to\infty$ or $0$.}. 
Thus, upon further compactification to $0-D$ on $T^2_T$, mirror symmetry 
implies that the partition function obeys $Z(S,U,T)=Z(U,S,T)$, as announced 
in eq.~(\ref{ZZ}) (for $T=i$, and after a Poisson resummation, the partition
function can be brought into the form (\ref{ZUS})).
The symmetry $U\leftrightarrow T$, leading to a triality, 
is manifest in the partition function,
since our compactification on $T^2_S\times T^2_U \times T^2_T$ is manifestly
symmetric in the interchange $T_U^2\leftrightarrow T^2_T$.
This triality of the (classical part) of the 1-loop partition function
$Z(S,U,T)$ of a $2-D$ sigma model on $T^2$ was observed in \cite{dvv}. In 
string theory, this triality is rather mysterious because $T$ is the complex 
structure of the world-sheet torus while $U$ and $S$ are the complex 
structure and K\" ahler structure of the target-space torus $T^2$.
However, for the 2-form theory on $T^2_S\times T^2_U \times T^2_T$ this
triality is the geometrical symmetry permuting the three two-tori. 

Two generalizations of our results are immediate.

First of all, the symmetry $S\leftrightarrow U$ holds for a
compactification $T^2_S \times T^2_U \times \Sigma$, with $\Sigma$ any
Riemann surface.

Secondly, the previous result can be further generalized to a generic
compactification $T^2_S\times \Sigma \times \tilde{\Sigma}$, with $\Sigma$
and $\tilde{\Sigma}$ any two Riemann surfaces, with the same choice for $V$
as in the previous examples. The compactification is manifestly invariant
under the interchange $\Sigma\leftrightarrow \tilde{\Sigma}$. By performing
the compactification in two stages, first from $6-D$ to $2-D$ on
$T^2_S\times \Sigma$, and then from $2-D$ to $0-D$ on $\tilde{\Sigma}$, the
previous symmetry gives rise to a symmetry under the interchange of
the world-sheet with the target space of a toroidal sigma model.
In detail, let us denote by $A_I$ and $B^I$, $I=1,..,g$, 
the 1-cycles of the Riemann surface $\Sigma$, of genus $g$.
The complex structure of the surface is determined by the data ${\rm G}$,
${\rm B}$:
\beq
\int_{\Sigma}B^I\wedge *B^J={\rm G}^{IJ},\;\; \int_{\Sigma}A_I\wedge*
B^J={\rm B}_{IL}{\rm G}^{LJ},\;\; \int_{\Sigma} A_I\wedge *A_J={\rm G}_{IJ} +
{\rm B}_{IL}{\rm G}^{LM}{\rm B}_{MJ}, \;\; {\rm B}_{IJ}={\rm B}_{JI}.
\eeq{m8}
We define the 2-cycles of $T^2_S\times\Sigma$ as
\ber
&& \a=a_S\wedge A_I, \;\; a_S\wedge B^I, \;\; a_S\wedge b_S, \nonumber \\
&& \b=B^I\wedge b_S,\;\; b_S\wedge A_I,\;\; \omega_{\Sigma}.
\eer{m9}
Here $\omega_\Sigma$ is the generator of $H_2(\Sigma,\Z)$. 
Following the same steps as before, we find a $2-D$ sigma model, propagating
on the world-sheet $\tilde{\Sigma}$, with constant background 
metric and antisymmetric tensor describing a torus $T^{2g}\times S^1$:
\beq
G_{ij}= S_2 \left(\begin{array}{ccc} {\rm G+ BG^{-1}B} & {\rm BG^{-1}} & 0
\\ {\rm G^{-1}B} & {\rm G^{-1}} & 0 \\ 0 & 0 & S_2^{-1} V_\Sigma/V_S 
\end{array}  \right), \;\; 
B_{ij} = S_1\left( \begin{array}{ccc} 0 & {\rm I} & 0 \\ -{\rm I} &0 &
0 \\ 0 & 0 & 0 \end{array} \right) ,
\eeq{m10}
where I is the $g\times g$ identity matrix and $V_\Sigma$ is the volume of 
$\Sigma$. In the two limits $V_S/V_\Sigma\rightarrow \infty$, 
$V_S/V_\Sigma \rightarrow 0$, the sigma model 
on $S^1$ decouples as remarked after equation~(\ref{m7}). 
Inserting the background (\ref{m10}) in the action (\ref{m5}) one finds that   
the ``left-over,'' i.e. a sigma model on $T^{2g}$ propagating on $\tilde\Sigma$,
is the same as a $2\tilde{g}$-dimensional toroidal sigma model  
propagating on the world-sheet $\Sigma$, with background given by 
eq.~(\ref{m10}) with ${\rm G}$, ${\rm B}$ replaced everywhere by the data
of $\tilde{\Sigma}$: $\tilde{\rm G}$ and $\tilde{\rm B}$. 
Notice that, when the genus of the two Riemann surfaces is different, 
$\tilde{g}\neq g$, and the two sigma models dual
to each other contain a different number of fields.

Other dualities of the classical partition function, interchanging the
world-sheet with the target space, were considered in ref.~\cite{gmr}. One of
them can be interpreted geometrically by slightly modifying the construction
presented above. We still compactify a $6-D$ two-form, with action given in
eq.~(\ref{m1}), on $T^2_S\times \Sigma \times \tilde{\Sigma}$, with two
changes.
First of all we choose the following two-cycle basis for $T^2_S\times
\Sigma$:
\ber
&& \a= a_S\wedge A_I,\;\; b_S\wedge A_I,\;\; a_S\wedge b_S , \nonumber \\ &&
\b=B^I\wedge b_S,\;\; a_S\wedge B^I,\;\; \omega_\Sigma.
\eer{m11}
Secondly, instead of eq.~(\ref{m3}), we impose
\beq
\int_\Sigma B^I\wedge (H-dB)=0,\;\; I=1,...,g.
\eeq{m12}
The configurations that minimize the classical action eq.~(\ref{m1}) are
harmonic (they obey $dH=0$, $d*H=0$), therefore, they 
can be expanded in the basis $\a,\b$:
\beq
H=\a_i\wedge \Pi^i + \b^i \wedge \Pi^D_i, \;\; B= \a_i \wedge X^i + \b^i \wedge
X^D_i.
\eeq{m13}
Equation~(\ref{m12}) implies $\Pi^i=dX^i$. Standard manipulations~\cite{v} lead
to the sigma-model action eq.~(\ref{m5}), describing the target space
$T^{2g}\times S^1$, with metric and
antisymmetric tensor given by
\beq
G_{ij}=\left(\begin{array}{cc} M(S)\otimes{\rm G} & 0 \\ 0 & V_\Sigma/V_S 
\end{array} \right), \;\;
B_{ij}=\left(\begin{array}{cc} \varepsilon\otimes {\rm B} & 0 \\ 0 & 0 
\end{array} \right),\;\;
\varepsilon=\left(\begin{array}{cc} 0 & 1 \\ -1 & 0 \end{array} \right) ,
\eeq{m14}
where ${\rm G}_{IJ}, {\rm B}_{IJ}$ are defined in (\ref{m8}) and $M(S)$ is
defined in (\ref{MS}).
When $V_\Sigma/V_S\rightarrow \infty,0$, the sigma model on $S^1$ decouples 
and one is left as before 
with a $2g$-dimensional sigma model on the toroidal background $T^{2g}$. 
Upon compactification to $0-D$ on $\tilde{\Sigma}$, the classical 
equations of motion of $H$ and the constraint~(\ref{m12}) imply
\footnotemark
\beq
\int_{\tilde{\Sigma}}\tilde{B}^I\wedge (H-dB)=0.
\eeq{m15}
\footnotetext{We implicitly assumed throughout the paper that $B$, upon
compactification on $M^6$, is
periodic up to integral elements of $H_3(M^6,\Z)$, i.e. $[dB]\in
H_3(M^6,\Z)$.}

The ``classical partition function,'' 
$Z({\rm G,B}, {\rm \tilde{G},\tilde{B}},S)$, 
is the sum over all solutions of the
classical equations of motion of $\exp(-S_{2-D}|_{class})$, where
$S_{2-D}|_{class}$ is the two-dimensional action evaluated at the classical
solution. {\em On the classical solutions} (but not in general!) our
dimensional reduction is symmetric in the interchange of the
target-space data ${\rm G,B}$ with the world-sheet data ${\rm
\tilde{G},\tilde{B}}$ (cfrs. eqs.~(\ref{m12},\ref{m15})); this can be made
explicit by
performing a Poisson resummation in the partition function. This symmetry is
one of the target space for world-sheet dualities observed (at $S=i$) in
ref.~\cite{gmr}.  
\noindent
\section{$(M^4,\t)\leftrightarrow (\tM^4,\tt)$ Duality and Compactifications of
a Self-Dual 4-Form Theory in $10-D$ }
\setcounter{equation}{0}
In ref.~\cite{v}, it was shown that by compactifying a $10-D$ self-dual
four-form, $B$, on $K^3\times T^2_S$, one obtains a $4-D$ Abelian gauge theory 
with group $U(1)^{b_2}$ (where $b_2$ is the second Betti number of $K^3$).
This result can be generalized to a compactification $M^4\times T^2_S$,
where $M^4$ is any smooth manifold with $b_1=0$.
Let us denote by $\gamma_I$, $I=1,..,b_2$, a basis for $H_2(M^4,\Z)$,    
and define
\beq
G_{IJ}=\int_{M^4}\gamma_I\wedge *\gamma_J,\;\;
Q_{IJ}=\int_{M^4}\gamma_I\wedge \gamma_J.
\eeq{m16}
When $b_1(M^4)=0$, the dimensional reduction of a five-form field strength 
$H$ reads 
\beq
H=\a_I\wedge F^I + \b^I\wedge F^D_I, \;\; B=\a_I\wedge A^I + \b^I\wedge
A^D_I, \;\; \a_I=\gamma_I\wedge a_S,\;\; \b^I=\gamma_I\wedge b_S.
\eeq{m17}
Choosing as usual a global vector field $V\in T^2_S$, such that $i_V\b^I=0$,
one finds a $4-D$ action for $b_2(M^4)$ Abelian gauge fields~\cite{v}
\beq
S_{4-D}={1\over g^2}\int F^I \wedge G_{IJ} *F^J -i{\theta\over 8\pi^2}\int
F^I\wedge Q_{IJ} F^J.
\eeq{m18}   
Recall that $S=\theta/2\pi + 4\pi i/g^2$.
Upon further compactification on $M^4\times T^2_S \times \tilde{M}^4$, one
finds a manifest symmetry under the interchange of $G,Q$ with
$\tilde{G},\tilde{Q}$. In other words, one finds a duality between the
geometrical data of the manifold $\tilde{M}^4$ and the coupling-constant
matrix $\tau_{IJ} =\theta Q_{IJ}/2\pi+4\pi i G_{IJ}/g^2$.

This symmetry, obvious in our construction, would look rather puzzling
from the four-dimensional point of view. Explicitly, in four dimensions, this
duality relates a theory on a space-time manifold $M^4$ and with a 
coupling-constant matrix $\tau$ to a theory on a {\em different} four-manifold 
and with a different gauge-coupling matrix:
\beq
\{M^4(G,Q), \tau(\tilde{G},\tilde{Q})\} \leftrightarrow
\{M^4(\tilde{G},\tilde{Q}), \tau(G,Q)\} .
\eeq{44}

The construction explained here sometimes works for more than the partition
function of an Abelian gauge theory. 

In particular, when both $M^4$ and $\tilde{M}^4$ are topologically $K^3$
surfaces, our construction can be embedded in an $N=4$ compactification of
the Type IIB string, since this string contains among its massless fields a
self-dual $10-D$ four-form. Our construction, in other words, computes the
large-volume limit $V_{M^4}\rightarrow\infty$ of the partition function of
a type IIB string on $K^3\times T^2_S\times \tilde{K}^3$. For small values
of the type IIB string coupling constant~\footnote{Equivalently, 
because of Type IIB strong-weak coupling duality, for large values of the 
coupling constant.} the gravitational sector of the theory
decouples, and the partition function becomes that of a rigid $N=4$
supersymmetric Yang-Mills theory compactified on $\tilde{K}^3$. This
partition function has been studied (for an $SU(2)$ gauge group) in
ref.~\cite{vw}. It would be interesting to see in detail how the space-time
$\leftrightarrow$ gauge-coupling duality works in this case.        

Finally, we should mention more duality symmetries of the classical
partition sum of free $U(1)^r$ gauge theories on four-manifolds $M^4$.
In \cite{v,w2}, it was shown that such partition functions are formally equal 
to those of a $2-D$ toroidal model with a genus-$r$ world-sheet, and with 
left-handed and right-handed $2-D$ momenta+windings in a self-dual Lorentzian
lattice $\Gamma^{b_+,b_-}$, with signature $(b_+,b_-)$, where $b_+$ ($b_-$) is
the number of (anti-)self dual harmonic two-forms in $M^4$. The data of $M^4$
are encoded in $\Gamma^{b_+,b_-}$, while the gauge-couplings data are encoded 
in the period matrix $\tau$ of the world-sheet ($\tau$ can be extended 
to a general positive-definite symmetric complex matrix). Such partition 
functions have many duality symmetries which mix the space-time data with the
world-sheet data, similar to the target-space $\leftrightarrow$
world-sheet dualities considered in \cite{gmr}. 
As discussed in this work, some of these dualities can be
interpreted as the geometrical symmetries of compactified theories in higher
dimensions. More dualities could have their geometrical origin in some simple
generalizations of the compactifications considered here. One such instance
is the compactification of a $10-D$ self-dual 4-form theory on 
$M^4\times \Sigma\times \tM^4$, with $\Sigma$ any Riemann surface (this has
$S$-duality only if $\Sigma =T^2$).

To recover all the symmetries of $4-D$ gauge 
theories, and their possible geometrical origin from higher dimensions, is an
interesting problem, 
which may shed more light on the non-perturbative dynamics 
of supersymmetric gauge theories and strings. In this paper, we considered
aspects of this problem in some simple, yet probably significant cases.

\vskip .3in \noindent
{\bf Acknowledgements} \vskip .2in \noindent
AG thanks the Department of Physics of NYU for its kind hospitality.
The work of AG is supported in part by BSF - American-Israel Bi-National
Science Foundation, and by the BRF - the Basic Research Foundation.
The work of MP is supported in part by NSF under grant PHY-9318781. 
 
\vskip .3in \noindent


\begin{thebibliography}{6666}
 
\newcommand{\np}{Nucl.\ Phys.\ }
\newcommand{\pr}{Phys.\ Rev.\ }
\newcommand{\cmp}{Commun.\ Math.\ Phys.\ }
\newcommand{\pl}{Phys.\ Lett.\ }

 
\bibitem{ggpz3}  L. Girardello, A. Giveon, M. Porrati and A. Zaffaroni, 
                 hep-th/9507064.
\bibitem{se} K. Intriligator and N. Seiberg, hep-th/9509066.
\bibitem{gpr} For a review, see A. Giveon, M. Porrati and E. Rabinovici,
              hep-th/9401139, Phys. Rep. {\bf 244} (1994) 77.
\bibitem{p} J. Polchinski, hep-th/9511157.
\bibitem{ggpz}  L. Girardello, A. Giveon, M. Porrati and A. Zaffaroni, 
                hep-th/9406128, Phys. Lett. {\bf B334} (1994) 331; 
                hep-th/9502057, Nucl. Phys. {\bf B448} (1995) 127.
\bibitem{w} E. Witten, hep-th/9507121.
\bibitem{s} J. Schwarz, hep-th/9604171.
\bibitem{v} E. Verlinde,  hep-th/9506011, Nucl. Phys. {\bf B455} (1995) 211.
\bibitem{ht} M. Henneaux and C. Teitelboim, Phys. Lett. {\bf B206} (1988) 650.
\bibitem{dvv} R. Dijkgraaf, E. Verlinde and H. Verlinde, in 
           {\em Perspectives in String Theory, proceedings} (Copenhagen, 1987)
            117.
\bibitem{gmr} A. Giveon, N. Malkin and E. Rabinovici, Phys. Lett. {\bf B220}
(1989) 551.
\bibitem{vw} C. Vafa and E. Witten, Nucl. Phys. {\bf B431} (1994) 3.
\bibitem{w2} E. Witten, hep-th/9505186. 
\end{thebibliography}
\end{document}